\documentstyle[12pt,aasms4]{article}

\newcommand{\lsun}{log $L/L_{\odot}\,$}
\newcommand{\msun}{$M/M_{\odot}\,$}

\begin{document}

\title{On the pulsation mode identification of short-period Galactic Cepheids}

\author{G. Bono\altaffilmark{1}, W.P. Gieren\altaffilmark{2}, 
M. Marconi\altaffilmark{3}, and P. Fouqu\'e\altaffilmark{4}}

\affil{1. Osservatorio Astronomico di Roma, Via Frascati 33,
00040 Monte Porzio Catone, Italy; Visiting Astronomer, ESO/Santiago, 
Chile; bono@coma.mporzio.astro.it}
\affil{2. Dept. de Fisica, Grupo de Astronomia, Univ. de Concepcion, 
Casilla 160-C, Concepcion, Chile; wgieren@coma.cfm.udec.cl}   
\affil{3. Osservatorio Astronomico di Capodimonte, Via Moiariello 16,
80131 Napoli, Italy; marcella@na.astro.it}
\affil{4. Observatoire de Paris-Meudon, DESPA F-92195 Meudon Cedex, France; 
and ESO, Casilla 19001, Santiago 19, Chile; pfouque@eso.org}

\begin{abstract}
We present new theoretical Period-Radius (PR) relations for first overtone 
Galactic Cepheids. Current predictions are based on several sequences of 
nonlinear, convective pulsation models at solar chemical composition 
(Y=0.28, Z=0.02) and stellar masses ranging from 3.0 to 5.5 $M_\odot$. 
The comparison between predicted and empirical radii of four short-period 
Galactic Cepheids suggests that QZ Nor and EV Sct are pulsating in the 
fundamental mode, whereas Polaris and SZ Tau pulsate in the first overtone. 
This finding supports the mode identifications that rely on the comparison 
between direct and Period-Luminosity (PL) based distance determinations 
but it is somewhat at variance with the mode identification based on 
Fourier parameters. In fact, we find from our models that fundamental and 
first overtone pulsators attain, for periods ranging from 2.7 to 4 d, 
quite similar $\phi_{21}$ values, making mode discrimination from this 
parameter difficult. The present mode identifications for our sample of 
Cepheids are strengthened by the accuracy of their empirical radius 
estimates, as well as by the evidence that predicted fundamental and first 
overtone radii do not show, within the current uncertainty on the 
Mass-Luminosity (ML) relation, any degeneracy in the same period range. 
Accurate radius determinations are therefore an excellent tool to 
unambiguously determine the pulsation modes of short-period Cepheids.
\end{abstract}


\pagebreak 

\section{Introduction} 

Classical Cepheids are an important link between stellar and extragalactic 
research, and they are widely adopted not only to estimate cosmic 
distances (Feast 1999; Gieren, Fouqu\'e, \& Storm 2000) but also to 
investigate young stellar populations in external galaxies 
(Magnier et al. 1997; Macri et al. 2001). Given the extreme usefulness 
of Cepheids for various astrophysical fields, it is crucial to establish 
their physical properties with high accuracy. Even though both 
evolutionary and pulsational properties of Cepheids are based on robust 
theoretical predictions we are still dealing with several long-standing 
unsolved problems, such as the universality of both PL 
and Period-Luminosity-Color (PLC) relations (Bono et al. 1999a; 
Gieren et al. 1999; Caputo et al. 2000; Groenewegen 2000), and the 
explanation of the Fourier parameters of the Bump Cepheids (Feuchtinger, 
Buchler, \& Kollath 2000). Moreover, the large photometric databases 
collected by the microlensing experiments (EROS, MACHO, OGLE)  
have provided the unique opportunity to investigate the occurrence  
of exotic objects in the Magellanic Clouds, such as mixed-mode Cepheids. 
At the same time, the excellent light curve phase coverage and the large 
sample of Cepheids measured by these projects gave also the opportunity 
to increase, by more than one order of magnitude, the number of detected 
first overtone (FO) Cepheids, and  to settle (Beaulieu et al. 1995; 
Welch et al. 1995) the long-standing problem (Pel \& Lub 1978; 
Gieren 1982; B\"ohm-Vitense 1988) concerning the occurrence of this mode 
among classical Cepheids. 

One of the main results of these investigations was that fundamental (F)  
and FO Cepheids are distributed, as expected, along two distinct 
sequences in the magnitude-period plane and present quite different Fourier 
parameters in certain period ranges. The latter is a crucial finding for two 
different reasons: 
i) classical Cepheids, at variance with RR Lyrae stars, cannot be easily 
split into F and FO variables according to the shape of their light curves. 
ii) the result strongly supports the evidence originally brought forward by 
Antonello, Poretti \& Reduzzi (1990), based on the light curve Fourier 
parameters, that the so-called s-Cepheids pulsate in the first overtone. 
This indication was further strengthened by the evidence that the Fourier 
parameters of Cepheid radial velocity curves do show a very similar behavior 
(Kienzle et al. 1999; Moskalik et al. 2001) as their light curve 
counterparts. However, even though the results disclosed by the massive 
microlensing data can hardly be questioned, it is worth mentioning that 
by comparing directly measured
distances based on the infrared surface brightness method and distance 
estimates based on the PL relation, Gieren, Fouqu\'e, 
\& Gomez (1997, 1998; hereinafter GFG97, GFG98) found evidence that at 
least some short-period, s-Cepheids could rather be F than FO pulsators.    
The overall scenario is jazzed up by the fact that current Cepheid models 
do not account for the Fourier parameters of 
observed light and radial velocity curves (Antonello \& Aikawa 1995; 
Feuchtinger et al. 2000). In view of these facts, it is desirable to 
have additional quantitative criteria which allow us a firm identification 
of Cepheid pulsation modes.

Parallel to the mentioned studies, several investigations in the recent 
literature have been devoted to the comparison between theoretically 
predicted, and 
empirical measurements of Galactic Cepheid radii (Laney \& Stobie 1995; 
Bono, Caputo, \& Marconi 1998, hereinafter BCM98; Ripepi et al. 1998; 
Gieren et al. 1999; Nordgren et al. 2000). Most of this work has
focused on Cepheids 
pulsating in the fundamental mode, since these variables are characterized 
by larger luminosity and radial velocity amplitudes when compared to first 
overtones which makes it easier to obtain accurate radii for them, 
particularly when Baade-Wesselink type techniques of radius determination 
are employed.
However, thanks to optical interferometric measurements Nordgren 
et al. (2000) recently succeeded in determining the mean angular 
diameter of the nearest Cepheid, $\alpha$ UMi (Polaris), from which they 
were able to estimate its radius. The pulsation behavior of this Cepheid 
is somewhat peculiar (Kamper \& Fernie 1998) but it has been recently 
classified as a first overtone (Feast \& Catchpole 1997). 
In order to compare the new radius evaluation with both empirical and 
theoretical PR relations, Nordgren et al. were forced to 
fundamentalise the period of Polaris, i.e. they added 0.148 to its 
logarithmic period, since at present we still lack both empirical and 
theoretical PR relations for first overtones.   

The main aim of this Letter is to present a new theoretical PR relation 
for FO Galactic Cepheids constructed by adopting a fixed chemical 
composition (Y=0.28, 
Z=0.02) and several stellar masses ranging from 3.0 to 5.5 $M_\odot$, 
and exploit this relation as a new tool for discriminating F and FOs
among classical Cepheids.
To account for current uncertainties on the predicted luminosity of 
intermediate-mass Cepheids, the models were constructed by adopting 
two different ML relations.  
In \S 2 we discuss current empirical evidence concerning 
the mode identification of short-period Cepheids. In \S 3 we compare 
the predicted PR relations with observed Cepheid radii. Conclusions 
about the mode identification from measured radii, and a few comments 
on the developments of this project close the paper.

\section{Empirical facts and theoretical predictions}
 
To assess the reliability of the mode identification for classical Cepheids 
based on their 
Fourier parameters, we selected four short-period Galactic Cepheids,
namely EV Sct, SZ Tau, QZ Nor, and Polaris as test objects.
These variables were selected because they have accurate, period-independent 
individual distance and radius estimates from the infrared surface brightness 
technique 
(GFG97,98) and trigonometric parallaxes (Nordgren et al. 2000), 
respectively, and all of them are generally considered to be FOs based 
on their short-periods and low-amplitude, nearly 
sinusoidal light curves. Also, they have  accurate spectroscopic and 
photometric data and Table 1 summarizes their key empirical observables.  

We already mentioned that Fourier parameters are widely adopted in the 
literature to identify the pulsation mode, and therefore we evaluated 
the Fourier parameters $R_{21}$ and $\phi_{21}$ for the selected Cepheids. 
We estimated these parameters from the observed radial velocity curves, 
to avoid 
possible subtle errors affecting the transformation of the theoretical 
light curves onto the observational plane. Fig. 1 shows from top to 
bottom the Fourier parameters $R_{21}$, $\phi_{21}$, and $A_1$ of the four 
Cepheids together with the Fourier parameters for FO and F Galactic 
Cepheids by Kienzle et al. (1999) and 
Moskalik et al. (2001). A glance at the data plotted in this figure 
shows quite clearly that the Fourier parameters of our sample are 
indeed typical for Cepheids classified as first overtones.  

An independent observable we can adopt to assess whether a star 
is pulsating in the F or in the FO mode is the stellar 
radius. As a matter of fact, a star pulsating in the FO  
is brighter and larger than a star pulsating with the same period in 
the fundamental mode. Owing to the lack of empirical PR relations for 
FO Cepheids, we decided to use theoretical radii predicted 
by nonlinear models to perform this test for our four stars. New PR relations 
for fundamental mode Cepheids based on nonlinear, convective models were 
recently provided by BCM98. Even though the idea 
to compare predicted and empirical radii by artificially decreasing the 
fundamental period of the current theoretical PR relation is viable, it is 
quite risky. The reason is twofold:
i) the width in temperature of the instability region where the F  
mode is unstable is on average larger than for the FO. 
As a consequence, the comparison between fundamentalised variables and 
truly fundamental PR relations can hardly be adopted to constrain the 
pulsation mode;   
ii) recent observational data (EROS, OGLE) on Magellanic Cepheids seem 
to support the evidence that the slope of the fundamental PL relation 
changes at very short periods (Bauer et al. 1999; Groenewegen 2000). 
Therefore the PR relations for F and FO Cepheids could have different 
slopes.  

To avoid these complications we have constructed several new sequences 
of nonlinear, convective models by adopting a chemical composition typical 
of solar neighborhood Cepheids, i.e. Y=0.28, Z=0.02. In order to assess 
on a quantitative basis the difference, if any, between F and FO  
Cepheid PR relations we adopted the same theoretical framework 
as adopted by BCM98 and by Bono, Marconi, \& Stellingwerf (1999b). 
The reader interested in the details of the input physics is referred 
to these papers. Moreover, to account for the difference in the luminosity 
of intermediate-mass Cepheids predicted by evolutionary models that include 
(noncanonical) or neglect (canonical) convective core-overshooting during 
hydrogen burning phases, we adopted, according to BCM98, two different 
ML relations. The mass/luminosity values adopted for canonical models are:
\msun-\lsun=3.5-2.51, 4.0-2.72, 4.5-2.90, 5.0-3.97, and 5.5-3.07; while for 
the noncanonical models are: 3.0-2.52, 4.0-2.97, 4.6-3.19, and 4.75-3.24. 
In the former set the effective temperature of stable FO models ranges 
from 5800 to 6500 K, and from 5650 to 6200 K in the latter one.   
On the basis of these calculations we derived the following canonical 
and noncanonical PR relations:  
\[ 
\log R = 1.250(\pm0.005) + 0.755(\pm0.007) \log P \;\;\;\;\;\; \sigma=0.005  
\]  
\[ 
\log R = 1.219(\pm0.004) + 0.737(\pm0.005) \log P \;\;\;\;\;\; \sigma=0.004 
\]  
where R is the radius (solar units), P the period (d), and  $\sigma$ is 
the standard deviation. The pulsation properties of these models will be 
discussed in a forthcoming paper. The previous linear regressions bring 
out that the intrinsic dispersion of the FO PR relations 
is a factor of four smaller than for the F PR relations 
(0.005 against 0.02) derived by BCM98. Note also that the slopes of 
FO relations are significantly steeper than the one for F pulsators
($\approx0.75$ against $\approx0.65$). This difference is caused 
by two different effects: the FO instability region is both 
systematically narrower, and bluer than the F one. 
This means that FOs are, at fixed period, systematically brighter 
than F pulsators. These findings provide a clear argument that 
one should use {\em pure} F and FO PR relations to constrain the 
pulsation mode.


Fig. 2 shows the comparison of predicted Fourier parameters for 
both F (filled triangles) and FO pulsators (open circles) with those 
of our selected short-period Cepheids. The outcome of the comparison 
between theory and observations is not very comfortable, and indeed 
current models seem to suggest that for periods ranging from 2.7 to 
4 d both F and FO Cepheids attain similar $\phi_{21}$ values. The velocity 
amplitude $A_1$ and $R_{21}$ seem to be less affected by this degeneracy 
problem, since F pulsators attain values that are, at fixed period, larger 
than FO ones. The sequences of F and FO pulsators 
we constructed are too coarse to quantitatively assess the nature of 
the $\phi_{21}$ jump located at  $P\approx4.2$ d (see Fig. 1). 
As a consequence, 
this finding is only a preliminary hint and more models are 
necessary to establish the extent of this degeneracy. Nevertheless 
it seems to be clear that, in the 3-4 d period range, the mode 
identification based on $\phi_{21}$ values might be quite uncertain.

To shed new light on this problem we decided to investigate whether 
empirical radius measurements can be adopted to disentangle the mode 
identification problem.  
Fig. 3 shows the comparison of theoretical PR relations for both 
FO  (solid and dotted lines) and F (BCM98, dashed and 
dashed-dotted lines) pulsators with empirical data for the selected 
objects. The data plotted in this figure disclose 
several interesting results. At odds with the mode identification 
based on the Fourier parameters, the comparison between predicted 
and observed radii yield strong evidence that both EV Sct and QZ Nor  
are fundamental pulsators. 
This result seems quite robust for three reasons: 
i) predicted F and FO radii are, in this period range, substantially 
different; 
ii) the empirical radii have quite small uncertainties (both random 
and systematic). 
iii) the predicted dependence on metallicity is marginal in this period 
range. In fact, a decrease in the metal content from Z=0.0.02 to Z=0.008
causes, according to BCM98, a decrease in $\log R$ of approximately 
0.02 dex. Moreover, FO radii predicted by Bono, Castellani, \& Marconi 
(2000) for Z=0.008 attain, at fixed period, values that are in very 
good agreement with FO radii at solar metallicity. Thus suggesting 
that FO radii for $0.008 \leq Z \leq 0.02$ marginally depend on 
metallicity.   

As far as Polaris is concerned, its radius provides evidence that this object 
is indeed pulsating in the first overtone, as suggested by Feast \& Catchpole 
(1997) on the grounds of the Hipparcos distance estimate, although this 
conclusion is somewhat weakened by the current observational uncertainties. 
A detailed analysis of both light and radial velocity curves 
(Kamper \& Fernie 1998) suggest that Polaris is currently undergoing a 
transient phase during which both pulsation amplitudes and period are 
rapidly changing. Therefore, further accurate parallax measurements are 
needed to improve the accuracy of its radius determination. 
The empirical radius determination for SZ Tau seems to suggest that this 
object is also pulsating in the first overtone mode. We note here that 
to avoid any systematic uncertainty in the radius determination of this 
star due to the known variation of its pulsation period (Szabados 1977), 
we adopted the radius value based on the J-K solution in GFG97, since 
J and K data were collected simultaneously and do not present a potential 
phase misalignment problem as do the noncontemporaneous V and K data. 

Summarizing, the comparison between theoretically predicted and measured 
radii is unveiling that among the selected short-period Cepheids two 
objects are almost certainly fundamental mode pulsators, in contrast to 
the mode identification based on their Fourier parameters.  
The data plotted in Fig. 3 show two further interesting results. 
In contrast with BCM we find that even at short-periods emprical radii
are in very good agreement with theoretical predictions. 
The fundamental mode PR relations present an intrinsic dispersion of 
the order of 0.02 dex. Therefore empirical radius measurements of F 
pulsators can be hardly adopted to assess the accuracy of current 
ML relations. On the other hand, the intrinsic dispersion 
of FO PR relations is of the order of only 0.005 dex. 
This suggests that accurate FO radius determinations can supply 
useful constraints on both evolutionary and pulsational models.

\section{Conclusions and final remarks}

From the current work we find evidence that a comparison of accurately 
measured empirical Cepheid radii to theoretical radii for F and FO pulsators 
is a very useful tool to identify the pulsation modes of classical Cepheids.
In fact, theoretical predictions based on nonlinear, convective 
models seem to suggest that F and FO radii are 
not affected by any degeneracy for periods ranging from 2.7 to 4 days. 
This is not true for the Fourier parameter $\phi_{21}$ of radial velocity 
curves. In fact, two short-period Galactic Cepheids, namely EV Sct, and 
QZ Nor which on the basis of their Fourier parameters should be classified 
as FO pulsators are F mode pulsators, as revealed by their radii. 
The degeneracy between F  
and FO Fourier parameters is supported by current theoretical 
predictions. Even though the sequences of models we constructed 
are relatively coarse in the mass step, the Fourier parameters for 
fundamental pulsators ($M=4.5\div5.0M\odot$) based on theoretical 
radial velocity curves attain values quite similar to FOs with 
stellar masses ranging from 4.6 to 5.5 $M_\odot$. The radial velocity 
amplitude and $R_{21}$ seem to be less affected by this degeneracy.  

On the other hand, the comparison between predicted and observed 
radii strongly suggests that both Polaris and SZ Tau are pulsating 
in the first overtone, supporting the mode identification suggested 
by Feast \& Catchpole (1997) and by Laney (1997), respectively. 
In this context it is worth mentioning that our new PR relations 
for FO Cepheids present an intrinsic dispersion that is approximately 
a factor of four smaller that that for F PR relations. 
Moreover and even more importantly, we find that the 
slope of the former ones is steeper than that for the latter ones, 
suggesting that the mode identification is more robust 
if based on {\em pure} F and FO PR relations.   
Unfortunately, the number of short-period Cepheids, particularly 
s-Cepheids, for which accurate radius measurements are currently 
available is limited to a handful of objects. 
It goes without saying that new and accurate multiband, near-infrared 
photometric data and radial velocity measurements for such objects 
would be crucial to quantitatively assess the accuracy of the mode 
identification based on the PR relations.   
Finally, we remark that FO PR relations are much more sensitive to the 
ML relation and marginally dependent on the metallicity adopted to 
construct the pulsation models. Accurate radius measurements for 
FO Cepheids could therefore supply useful hints on this key relationship 
predicted by evolutionary models.

\acknowledgments
We would like to thank P. Moskalik for sending us his Fourier parameters 
for Galactic Cepheids in advance of publication and an anonymous referee 
for his/her pertinent suggestions.  
GB \& MM acknowledge financial support by MURST-Cofin 2000, under the 
project "Stellar observables of cosmological relevance".
WPG gratefully acknowledges partial financial support by Fondecyt projects 1000330 
and 8000002.

\pagebreak

\pagebreak
\begin{center}
\begin{tabular}{lcccc}
\tablewidth{0pt}\\
\multicolumn{4}{c}{TABLE 1. empirical data$^a$}\\
\hline
                     & EV Sct       & SZ Tau      &  QZ Nor     & $\alpha$ UMi\\
\hline
Period (d)           &3.09097       &3.148727     & 3.78673     & 3.97267  \\  
Radius ($R/R_\odot$) &$32.5\pm0.5$  &$45.6\pm4.0$ &$38.5\pm0.5$ & $46\pm3$ \\  
Distance (pc)        &$1635\pm25$   & $692\pm61$  &$1656\pm8$ &$132^{+9}_{-8}$\\
$R_{21}$ (RV)        &$0.19\pm0.05$ &$0.25\pm0.03$&$0.16\pm0.02$&$0.03\pm0.02$\\
$\phi_{21}$ (RV)     &$3.16\pm0.37$ &$3.48\pm0.13$&$3.70\pm0.15$&$3.90\pm0.38$\\
$A_1$ (RV)           &$7.74\pm0.27$ &$9.72\pm0.17$&$7.38\pm0.15$&$0.80\pm0.08$\\
\hline
\end{tabular}
\end{center}
\begin{minipage}{1.00\linewidth}
\noindent $^a$ Fourier parameters for EV Sct and SZ Tau were estimated 
according to data collected by Metzger et al. (1991) and Bersier et al. (1994), 
while for QZ Nor by Kienzle et al. (1999). Data for $\alpha$ UMi come from 
Hatzes \& Cochran (2000, period), Kamper \& Fernie (1998, $A_1$), 
Moskalik \& Oglozova (2000, $A_2$, $\phi_{21}$), and 
Nordgren et al. (2000, radius and distance). 
\end{minipage}

\pagebreak
\figcaption{
Top panel: amplitude ratio $R_{21}\equiv A_2/A_1$, i.e. the ratio between
second and first harmonic amplitude, as a function of period for the sample
of fundamental (triangles) and first overtone (open circles) Galactic
Cepheids collected by Moskalik et al. (2001), and Kienzle et al. (1999).
Data refer to the Fourier fit to the radial velocity curves. The four 
selected Cepheids were plotted by adopting different symbols.
Middle panel: same as the top panel, but for the phase difference
$\phi_{21}\equiv \phi_2-2\phi_1$, i.e. the phase difference between
second and first harmonic. Bottom panel: same as the top panel, but
for the first harmonic amplitude $A_1$. The error bars refer to internal
uncertainty in the Fourier fit.}   

\figcaption{Same as Fig. 1, but symbols refer to Fourier parameters of 
F and FO predicted radial velocity curves. The error bars show the 
internal uncertainty in the Fourier fit.}  

\figcaption{Comparison between empirical radius determinations and theoretically
predicted PR relations at solar chemical composition (Y=0.28, Z=0.02). 
Canonical and noncanonical PR relations for first overtone and fundamental
pulsators are plotted using different line styles. Symbols are the same as 
in Fig. 1.} 

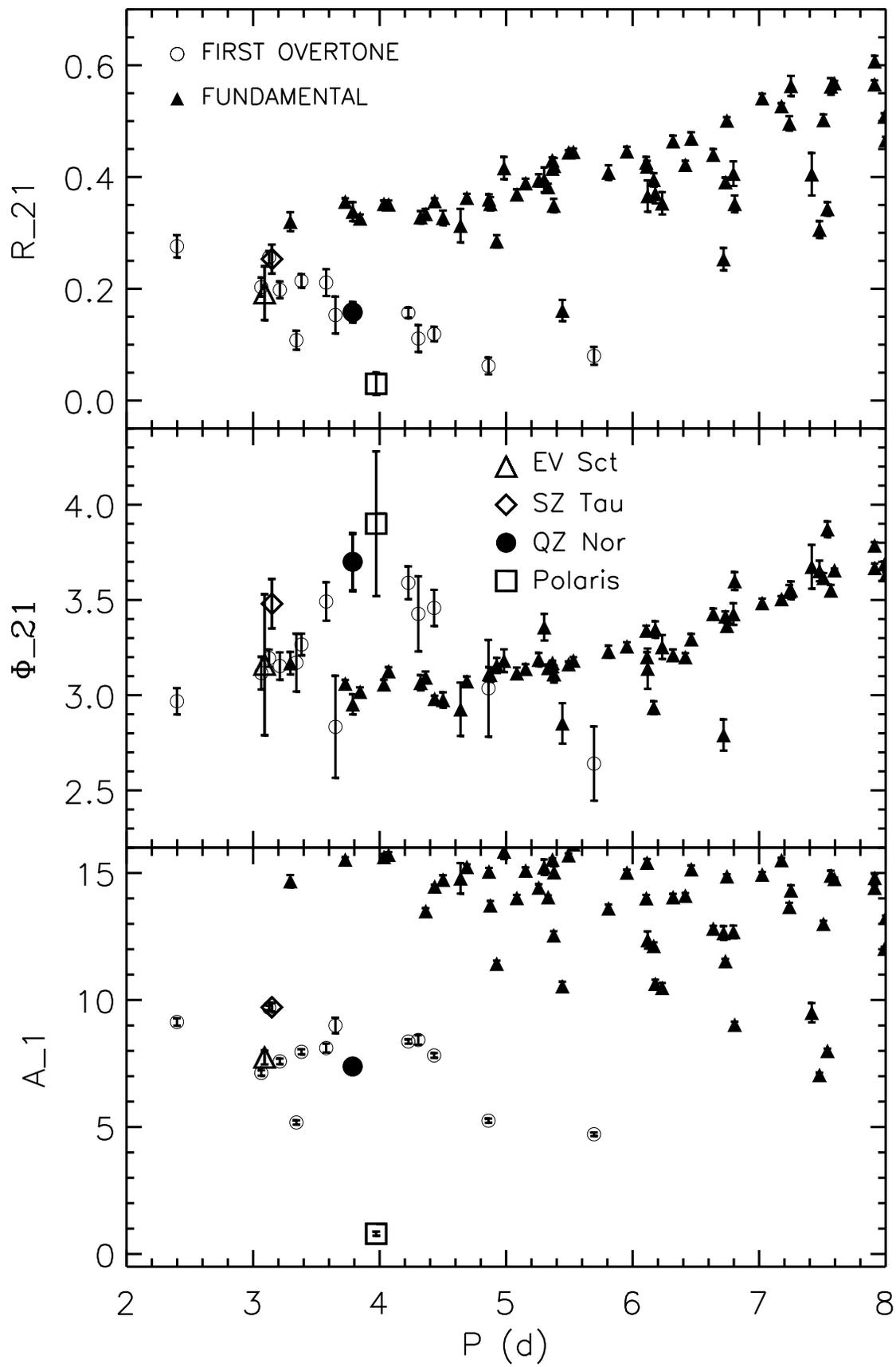
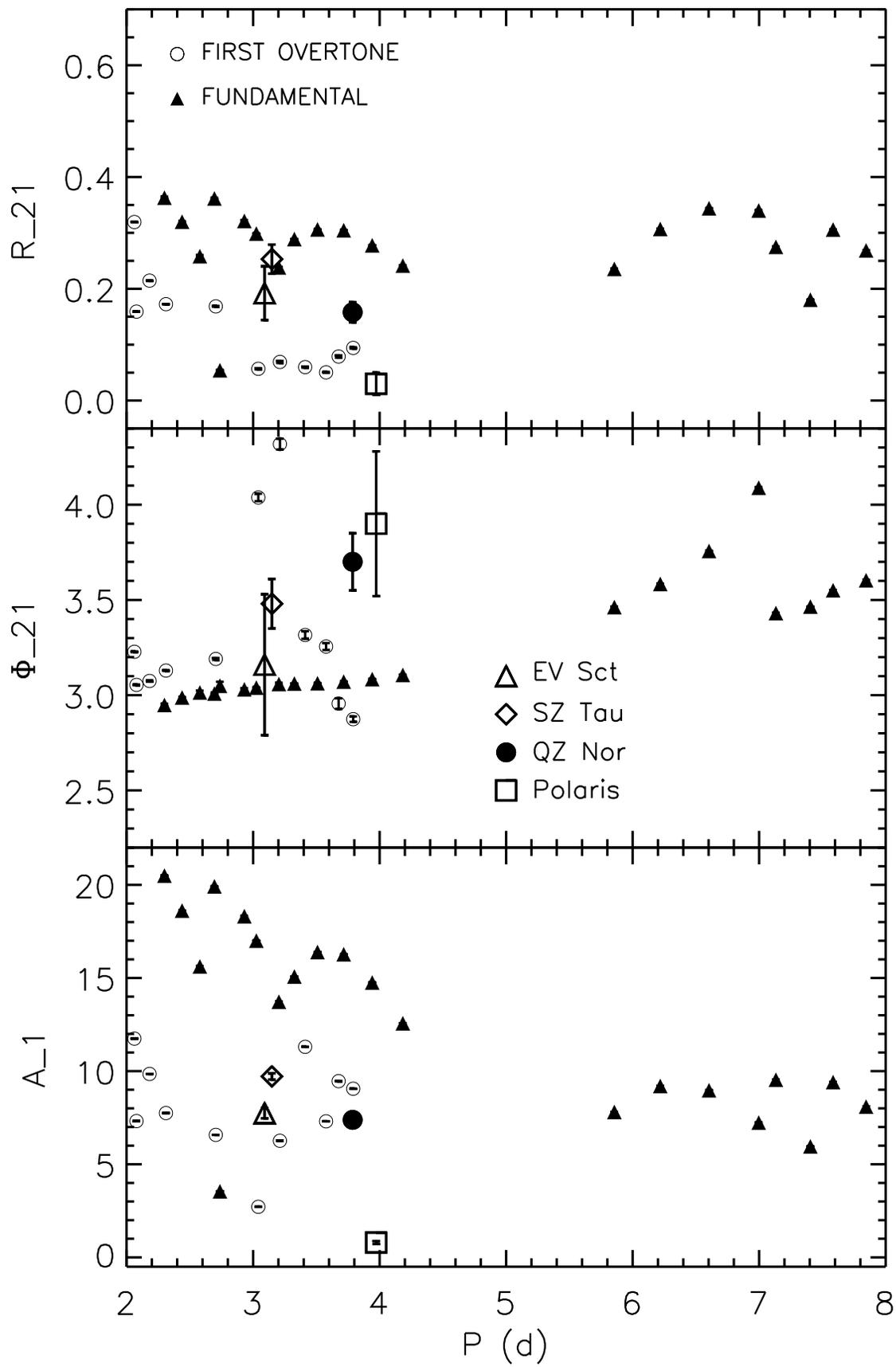
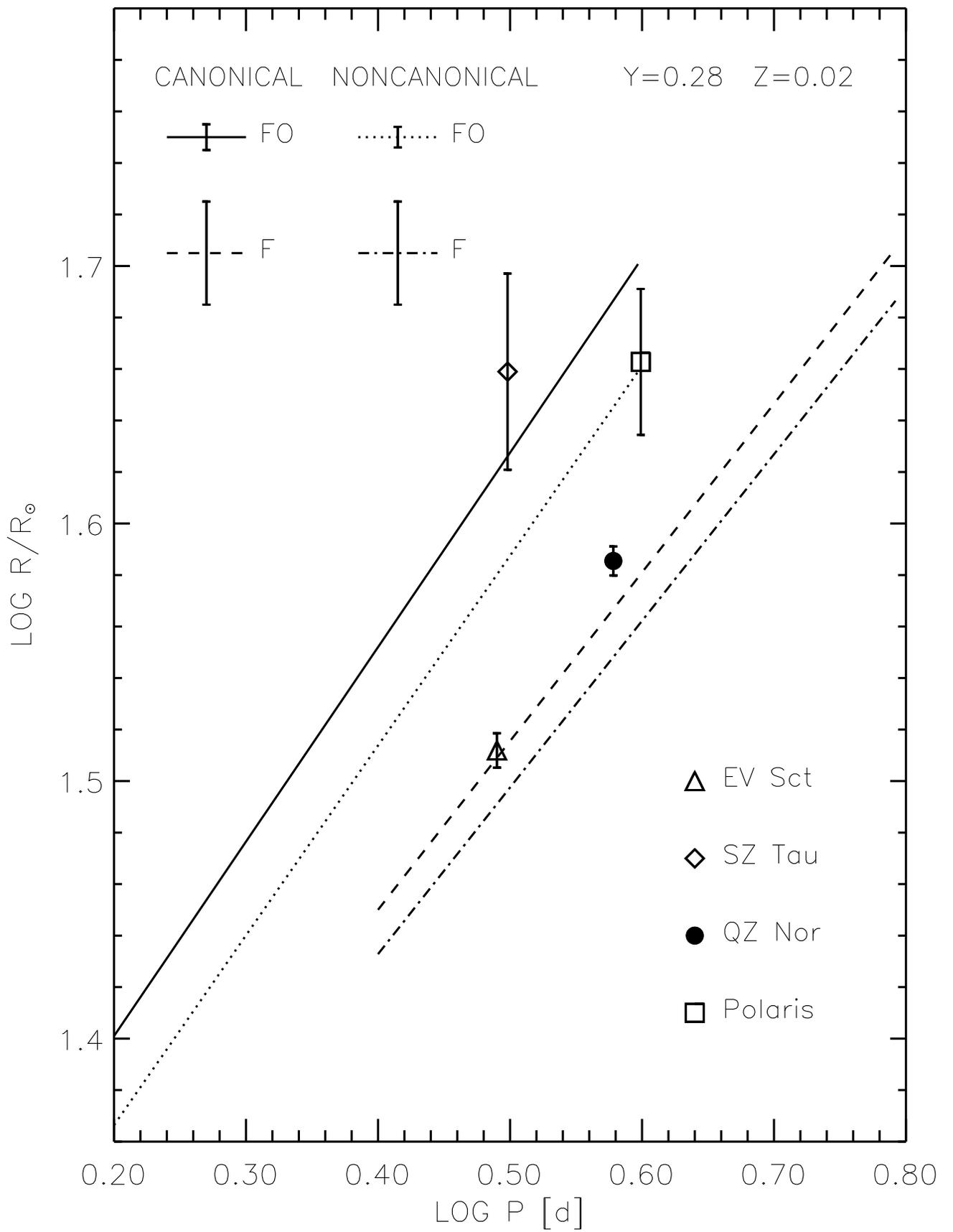
\end{document}